\documentclass[aps,prd,reprint,groupedaddress,nofootinbib,floatfix]{revtex4-2}

\usepackage{amssymb}
\usepackage{amsthm}
\usepackage{mathtools}
\usepackage{enumitem}
\usepackage{graphicx}
\usepackage{hyperref}
\usepackage{cleveref}
\usepackage[dvipsnames]{xcolor}
\usepackage{siunitx}
\usepackage{float}
\usepackage[T1]{fontenc}
\usepackage{booktabs}
\usepackage{natbib}

\hypersetup{
	colorlinks=true,
	linkcolor=blue,  
	urlcolor=blue,
	citecolor=blue
}

\newtheorem*{theorem}{Generalized multipole conjecture}

\begin{document}


\title{Multipole moments of a charged rotating disc of dust in general relativity}


\author{David Rumler}
\email[]{david.rumler@uni-jena.de}
\affiliation{Theoretisch-Physikalisches Institut, Friedrich-Schiller-Universität Jena, Max-Wien-Platz 1, D-07743 Jena, Germany}

\author{Reinhard Meinel}
\affiliation{Theoretisch-Physikalisches Institut, Friedrich-Schiller-Universität Jena, Max-Wien-Platz 1, D-07743 Jena, Germany}


\date{\today}

\begin{abstract}
	The gravitational and electromagnetic multipole moments of the charged rotating disc of dust, which is an axisymmetric, stationary solution of the Einstein-Maxwell equations in terms of a post-Newtonian expansion, are calculated and discussed. It turns out that the individual mass, angular momentum, electric and magnetic moments are ordered in the sense that higher moments have a lower absolute value. There is an interesting conjecture stating that the absolute values of all higher multipole moments of a uniformly rotating perfect fluid body are always greater than those of the corresponding Kerr spacetime, which we generalize to include charged bodies. We find that for the charged rotating disc of dust the conjecture holds (within the limits of accuracy of the post-Newtonian expansion).
\end{abstract}


\maketitle

\section{Introduction}
\label{sec:introduction}

The exterior spacetime of an astrophysical body can be characterized by its multipole moments. In 1970, Geroch introduced multipole moments in a coordinate independent way for static, asymptotically flat, vacuum spacetimes \cite{Geroch} and Hansen later generalized them to stationary (asymptotically flat, vacuum) spacetimes \cite{Hansen}. In case of axisymmetry, Hansen further showed that the multipole moment tensors can be represented by scalars $P_{n}$ \cite{Hansen}. Relying on an asymptotic expansion of the Ernst potential on the symmetry axis, Fodor, Hoenselaers and Perj\'{e}s presented, in 1989, an explicit algorithm to compute those scalar multipole moments \cite{10.1063/1.528551}.

A generalization to stationary, electro-vacuum spacetimes with gravitational and electromagnetic multipole moments, $P_{n}$ and $Q_{n}$, respectively, was published by Simon \cite{Simon} and a corresponding calculation scheme, in the axisymmetric case, was given by Hoenselaers and Perj\'{e}s \cite{Hoenselaers_1990}. As it turned out, \cite{Hoenselaers_1990} contained two non-trivial mistakes, however. The first one was corrected by Sotiriou and Apostolatos \cite{Sotiriou_2004} and the second one recently, in 2021, by Fodor et al.\ \cite{fodor}.

In the present paper we calculate and discuss the gravitational and electromagnetic multipole moments of the charged rotating disc of dust (with constant angular velocity and constant specific charge) using the latest calculation procedure by Fodor et al.\ \cite{fodor}. The charged rotating disc of dust is an axisymmetric, stationary and physically reasonable solution of the Einstein-Maxwell equations expressed in terms of a post-Newtonian expansion up to tenth order \cite{Palenta_2013,Breithaupt_2015,PhysRevD.94.104035}. An exact solution to this problem is not available so far; in particular it is not contained in the solution classes discussed in \cite{Klein2003}. A study of the multipole moments of the exact solution of the uncharged disc of dust \cite{Neugebauer:1993ct, PhysRevLett.75.3046, RFE} was published by Kleinwächter, Meinel and Neugebauer \cite{Kleinwächter}.

We additionally formulate a generalized version of the multipole conjecture by Filter and Kleinwächter \cite{Filter} and test it using the multipole moments of the charged rotating disc of dust. The conjecture states that the absolute values of all higher multipole moments of an axisymmetric, stationary, physically well-defined body are always greater than those of the corresponding multipole moments of the Kerr-Newman spacetime with the same mass, angular momentum and charge.
We show that the generalized multipole conjecture holds for the charged rotating disc of dust. 

\section{Charged rotating disc of dust}
\label{sec:disc}

In general relativity it is rather easy to construct purely mathematical solutions without any physical meaning, however, the charged rotating disc of dust is also physically relevant. We model the disc by an infinitesimally thin equilibrium configuration of dust with constant specific charge $\epsilon \in \left[-1,1\right]$ (electric charge density over baryonic mass density) that is rigidly rotating  around the axis of symmetry with constant angular velocity $\Omega$. Dust is a perfect fluid with vanishing pressure. The disc is axisymmetric, stationary and it obeys reflection symmetry \cite{Palenta_2013, Breithaupt_2015, Meinel_2015}. 

Using axisymmetry and stationarity the corresponding metric can be expressed globally in terms of Weyl-Lewis-Papapetrou coordinates\footnote{We use units in which $c=G=4\pi\epsilon_{0}=1$.}:
\begin{equation}\label{eq:bvp.4}
	\mathrm{d}s^2 = f^{-1}\left[ h\left( \mathrm{d}\rho^2 + \mathrm{d}\zeta^2 \right) + \rho^2\mathrm{d}\varphi^2 \right] - f\left( \mathrm{d}t + a\, \mathrm{d}\varphi \right)^2 \,.
\end{equation}
The electromagnetic four-potential $A_a$ takes the form:
\begin{equation}
	A_a = (0,0,A_{\varphi}, A_t)\,.
\end{equation}

In axistationary spacetimes the coupled Einstein-Maxwell equations in electro-vacuum can be reduced to the Ernst equations \cite{PhysRev.168.1415}:
\begin{align}
	\label{eq:bvp.1}
	\left( \Re \mathcal{E} + \lvert \Phi \rvert^2 \right) \Delta \mathcal{E} &= \left( \nabla\mathcal{E} + 2\bar{\Phi}\nabla\Phi \right) \cdot \nabla\mathcal{E} \,, \\
	\label{eq:bvp.2}
	\left( \Re \mathcal{E} + \lvert \Phi \rvert^2 \right) \Delta \Phi &= \left( \nabla\mathcal{E} + 2\bar{\Phi}\nabla\Phi \right) \cdot \nabla\Phi \,,
\end{align}
where 
\begin{equation}
	\mathcal{E} = \left(f-\lvert\Phi\rvert^2\right) + \mathrm{i}b \quad \text{and} \quad \Phi = \alpha + \mathrm{i}\beta
\end{equation}
are the Ernst potentials. Here $\alpha\coloneqq \Re\Phi=-A_t$ and
the potentials $\beta$ and $b$ are defined by
\begin{equation}\label{eq:bvp.6a}
	\beta_{,\rho} = \frac{f}{\rho}\left( A_{\varphi,\zeta} - aA_{t,\zeta} \right) \,, \quad \beta_{,\zeta} = -\frac{f}{\rho}\left( A_{\varphi,\rho} - aA_{t,\rho} \right)
\end{equation}
and
\begin{align}\label{eq:bvp.6b}
	b_{,\rho} &= - \frac{f^2}{\rho}a_{,\zeta} - 2\left( \beta A_{t,\rho}-A_t\beta_{,\rho} \right) \,, \notag \\ 
	 b_{,\zeta} &= \frac{f^2}{\rho}a_{,\rho} - 2\left( \beta A_{t,\zeta}-A_t\beta_{,\zeta} \right) \,.
\end{align}

With the help of reflection symmetry one can formulate a well-defined boundary value problem to the Ernst equations for the charged rotating disc of dust which was solved in terms of a post-Newtonian expansion up to eighth order by Palenta and Meinel \cite{Palenta_2013} and up to tenth by Breithaupt et al.\ \cite{Breithaupt_2015}:
\begin{gather}
	\label{eq:post-newton-expansion}
	f = 1 + \sum_{k=1}^{10}f_{2k}g^{2k} \,, \quad b = \sum_{k=1}^{10}b_{2k+1}g^{2k+1} \,, \\
	\label{eq:post-newton-expansion2}
	\alpha = \sum_{k=1}^{10}\alpha_{2k}g^{2k} \,, \quad \beta = \sum_{k=1}^{10}\beta_{2k+1}g^{2k+1} \,.
\end{gather}
The coefficient functions $f_{2k}$, $b_{2k+1}$, $\alpha_{2k}$ and $\beta_{2k+1}$ are functions of the elliptic coordinates $\eta\in[-1,1]$ and \mbox{$\nu\in[0,\infty]$}, defined via
\begin{equation}
	\rho = \rho_0\sqrt{\left(1-\eta^2\right)(1+\nu^2)} \,, \quad \zeta=\rho_0\eta\nu \,.
\end{equation}
Here, $g\coloneqq\sqrt{\gamma}$ and $\gamma$ is the relativity parameter, originally introduced by Bardeen and Wagoner \cite{bardeen}, defined by:
\begin{equation}
	\gamma \coloneqq 1 - \sqrt{f_{c}} \,, \quad \text{with} \quad f_{c} \coloneqq f\left(\rho=0, \zeta=0\right) \,.
\end{equation}
Note that the metric function $f$ can also be written as $f=e^{2U}$, where $U$ can be interpreted as a generalized Newtonian potential.
Equivalently, the relativity parameter can be expressed in terms of the redshift $Z_{c}$ of a photon travelling from the centre to infinity:
\begin{equation}
	\gamma = \frac{Z_{c}}{1+Z_{c}} \,.
\end{equation}

The parameter space of the disc solution is therefore given by $g\in[0,1]$, $\epsilon \in [0,1]$ and the coordinate radius $\rho_{0}$. For $g\ll1$ one obtains a Newtonian solution and $g \to 1$ corresponds to the ultra-relativistic limit in which we assume black hole formation. Without loss of generality we restrict to positive charges and $\rho_{0}$ serves as a scaling parameter. All multipole moments defined in the subsequent section are functions of $g$, $\epsilon$ and $\rho_{0}$ only.

\section{Multipole moments}
\label{sec:moments}

In order to derive multipole moments, we introduce new potentials
\begin{equation}\label{eq:new_potentials}
	\Xi \coloneqq \frac{1-\mathcal{E}}{1+\mathcal{E}} \quad \text{and} \quad \Lambda \coloneqq \frac{2\Phi}{1+\mathcal{E}} \,.
\end{equation}
Those can be expressed as an asymptotic expansion on the upper symmetry axis:
\begin{equation}
	\Xi_{+} = \frac{1}{\zeta}\sum_{n=0}^{\infty}\frac{m_{n}}{\zeta^n} \,, \quad \Lambda_{+} = \frac{1}{\zeta}\sum_{n=0}^{\infty}\frac{q_{n}}{\zeta^n} \,.
\end{equation}
According to Fodor  et al.\ \cite{fodor} the gravitational and electromagnetic multipole moments, $P_{n}$ and $Q_{n}$, respectively, of an axisymmetric and stationary spacetime can be obtained from the coefficients $m_{n}$ and $q_{n}$ by the following procedure:
\begin{align}
	P_{0} &= m_{0} \,, \\
	P_{1} &= m_{1} \,, \\
	P_{2} &= m_{2} \,, \\
	P_{3} &= m_{3} + \frac{1}{5}\bar{q}_{0}S_{10} \,, \\
	P_{4} &= m_{4} - \frac{1}{7}\bar{m}_{0}M_{20} + \frac{3}{35}\bar{q}_{1}S_{10} + \frac{1}{7}\bar{q}_{0}\left(3S_{20} - 2H_{20}\right) \,, \\
	P_{5} &= m_{5} - \frac{1}{21}\bar{m}_{1}M_{20} - \frac{1}{3}\bar{m}_{0}M_{30} + \frac{1}{21}\bar{q}_{2}S_{10} \notag \\ 
	&\quad+ \frac{1}{21}\bar{q}_{1}\left(4S_{20} - 3H_{20}\right) + \frac{1}{21}\bar{q}_{0}\left(\bar{q}_{0}q_{0}S_{10} \right. \notag \\
	&\quad\left.-\hspace{0.075cm}\bar{m}_{0}m_{0}S\!_{10} + 14S_{30} + 13S\!_{21} - 7H_{30}\right)
\end{align}
and
\begin{align}
	Q_{0} &= q_{0} \,, \\
	Q_{1} &= q_{1} \,, \\
	Q_{2} &= q_{2} \,, \\
	Q_{3} &= q_{3} - \frac{1}{5}\bar{m}_{0}H_{10} \,, \\
	Q_{4} &= q_{4} + \frac{1}{7}\bar{q}_{0}Q_{20} - \frac{3}{35}\bar{m}_{1}H_{10} - \frac{1}{7}\bar{m}_{0}\left(3H_{20} - 2S_{20}\right) \,, \\
	Q_{5} &= q_{5} + \frac{1}{21}\bar{q}_{1}Q_{20} + \frac{1}{3}\bar{q}_{0}Q_{30} - \frac{1}{21}\bar{m}_{2}H_{10} \notag \\ 
	&\quad- \frac{1}{21}\bar{m}_{1}\left(4H_{20} - 3S\!_{20}\right) + \frac{1}{21}\bar{m}_{0}\left(\bar{m}_{0}m_{0}H_{10} \right. \notag \\
	&\quad\left. -\hspace{0.075cm}\bar{q}_{0}q_{0}H_{10}-14H_{30} - 13H_{21} + 7S_{30}\right) \,,
\end{align}
where
\begin{gather}
	M_{ij} = m_{i}m_{j} - m_{i-1}m_{j+1} \,, \quad Q_{ij} = q_{i}q_{j} - q_{i-1}q_{j+1} \,, \\
	S_{ij} = m_{i}q_{j} - m_{i-1}q_{j+1} \,, \quad H_{ij} = q_{i}m_{j} - q_{i-1}m_{j+1} \,.
\end{gather}
Higher multipole moments become increasingly complicated. The gravitational and electromagnetic multipole moments are closely related to each other. In fact, by interchanging $m_{n} \leftrightarrow q_{n}$ and $\bar{m}_{n} \leftrightarrow -\bar{q}_{n}$ (correspondingly $M_{ij} \leftrightarrow Q_{ij}$ and $S_{ij} \leftrightarrow H_{ij}$), $P_{n}$  transforms into $Q_{n}$.

As mentioned in \cref{sec:disc} the disc of dust additionally obeys reflection symmetry. Expressed on the upper symmetry axis, reflection symmetry is equivalent to \cite{PanagiotisKordas_1995,RMeinel_1995,Pachón_2006,Ernst_2006,Meinel_2012}
\begin{equation}
	\mathcal{E}_{+}\left(\zeta\right)\bar{\mathcal{E}}_{+}\left(-\zeta\right) = 1\,, \quad \Phi_{+}\left(\zeta\right) = -\bar\Phi_{+}\left(-\zeta\right)\mathcal{E}_{+}\left(\zeta\right)
\end{equation}
or in terms of the new potentials to
\begin{equation}
	\Xi_{+}\left(\zeta\right) = - \bar{\Xi}_{+}\left(-\zeta\right) \,, \quad \Lambda_{+}\left(\zeta\right) = - \bar{\Lambda}_{+}\left(-\zeta\right) \,.
\end{equation}
This means that $P_{n}$ and $Q_{n}$ are real for even $n$ and imaginary for odd $n$.
The real and imaginary parts of $P_{n}$ are called mass and angular momentum moments
and those of $Q_{n}$ are referred to as electric and magnetic moments.
We denote:
\begin{align}
	P_{n} &= M_{n} + \mathrm{i}J_{n} \,, \\
	Q_{n} &= E_{n} + \mathrm{i}B_{n} \,,
\end{align}
where due to reflection symmetry,
\begin{align}
	&\text{for even $n$:} \quad P_{n} = M_{n} \,, \quad Q_{n} = E_{n} \,, \\
	&\text{for odd $n$:} \quad P_{n} = \mathrm{i}J_{n} \,, \quad Q_{n} = \mathrm{i}B_{n} \,.
\end{align}

Note that on the upper symmetry axis $\eta=1$, $\zeta=\rho_{0}\nu$ and thus
\begin{equation}
	\Xi_{+} = \frac{1}{\nu}\sum_{n=0}^{\infty}\frac{m_{n}^{\star}}{\nu^n} \,, \quad \Lambda_{+} = \frac{1}{\nu}\sum_{n=0}^{\infty}\frac{q_{n}^{\star}}{\nu^n}
\end{equation}
with dimensionless $m_{n}^{\star} \coloneqq \frac{m_{n}}{\rho_{0}^{n+1}}$ and  $q_{n}^{\star} \coloneqq \frac{q_{n}}{\rho_{0}^{n+1}}$.
In order to obtain coordinate independent expressions for the multipole moments, we normalize $m_{n}$ and $q_{n}$ by the disc's proper radius $R_{0} \coloneqq \int_{0}^{\rho_{0}}\!\sqrt{g_{\rho\rho}}\,\mathrm{d}\rho$, where $g_{\rho\rho}=f^{-1}h$, see also \cite{Rumler}: 
\begin{equation}
	m_{n}^{\circ} \coloneqq \frac{m_{n}}{R_{0}^{n+1}} \,, \quad q_{n}^{\circ}\coloneqq \frac{q_{n}}{R_{0}^{n+1}} \,.
\end{equation}

The first multipole moments are the gravitational mass, \mbox{$P_{0}=M_{0}=M$}, the angular momentum, \mbox{$P_{1}/\mathrm{i}=J_{1}=J$}, the electric charge, $Q_{0}=E_{0}=Q$, and the magnetic dipole moment, $Q_{1}/\mathrm{i}=B_{1}=D$. 

Inserting the potentials $f$, $b$, $\alpha$ and $\beta$, \cref{eq:post-newton-expansion,eq:post-newton-expansion2}, into $\Xi$ and $\Lambda$, \cref{eq:new_potentials}, and applying the above mentioned procedure reveals the multipole moments of the charged rotating disc of dust in terms of a post-Newtonian expansions up to tenth order.
The first multipole moments, normalized by the disc's proper radius $R_{0}$, up to third order ($k=3$), read\footnote{Note that the notation with the circle superscript introduced here, i.e.\ $P_{n}^{\circ} \coloneqq \frac{P_{n}}{R_{0}^{n+1}}$ and $Q_{n}^{\circ} \coloneqq \frac{Q_{n}}{R_{0}^{n+1}}$, has a different meaning than that in \cite{Breithaupt_2015}.}:
\begin{align}
	M^\circ &= \frac{4 g^{2}}{3 \pi}-\frac{4 \left(\epsilon^{2}-1\right) g^{4}}{45 \pi} \notag \\
	&\quad+\frac{1}{30240 \pi^{3}}\left(\left(2790 \epsilon^{4}-23699 \epsilon^{2}+20464\right) \pi^{2}\right. \notag \\ 
	&\quad\left.-\,35840 \epsilon^{4}+247296 \epsilon^{2}-211456\right) g^{6}
	+ \mathcal{O}\left(g^8\right) \,,
\end{align}
\begin{align}
	J^\circ &= \sqrt{1-\epsilon^{2}}\, \left[\frac{8 g^{3}}{15 \pi}-\frac{2 \left(34\epsilon^{2}-7\right) g^{5}}{315 \pi}\right. \notag \\
	&\quad+\frac{1}{453600 \pi^{3}}\left(\left(59894 \epsilon^{4}-40131 \epsilon^{2}-126312\right) \pi^{2}\right. \notag \\
	&\quad\left.\left.-\,322560 \epsilon^{4}+207872 \epsilon^{2}+1189888\right) g^{7}
	+ \mathcal{O}\left(g^9\right) \vphantom{ \frac{2 \left(34\epsilon^{2}-7\right) g^{5}}{315 \pi} }\right] \,,
\end{align}
\begin{align}
	Q^\circ &= \epsilon\, \left[\frac{4 g^{2}}{3 \pi}-\frac{16\left(\epsilon^2 -1\right) g^{4}}{45 \pi}\right. \notag \\ 
	&\quad+\frac{1}{30240 \pi^{3}}\left(\left(5862 \epsilon^{4}-32147 \epsilon^{2}+25840 \right) \pi^{2}\right. \notag \\
	&\quad\left.\left.-\,35840 \epsilon^{4}+247296 \epsilon^{2}-211456 \right) g^{6}
	+ \mathcal{O}\left(g^8\right) \vphantom{ \frac{16\left(\epsilon^2 -1\right) g^{4}}{45 \pi} }\right] \,,
\end{align}
\begin{align}
	D^{\circ} &=  \epsilon \sqrt{1-\epsilon^{2}}\, \left[\frac{4 g^{3}}{15 \pi}-\frac{\left(34\epsilon^{2}-115\right) g^{5}}{315 \pi}\right. \notag \\
	&\quad+\frac{1}{907200 \pi^{3}}\left(\left(58614 \epsilon^{4}-177091 \epsilon^{2}-200232 \right) \pi^{2}\right. \notag \\
	&\quad\left.\left.-\,322560 \epsilon^{4}-867328 \epsilon^{2}+3340288 \right) g^{7}
	+ \mathcal{O}\left(g^9\right) \vphantom{ \frac{\left(34\epsilon^{2}-115\right) g^{5}}{315 \pi} }\right] \,.
\end{align}
Below, we also list the normalized gravitational and electromagnetic quadrupole and octupole moments (up to $k=3$) of the charged rotating disc of dust:
\begin{align}
	P_{2}^{\circ} &= M_{2}^{\circ} \notag \\ 
	&= -\frac{4 g^{2}}{15 \pi}+\frac{4 \left(5\epsilon^{2}+8\right) g^{4}}{105 \pi} \notag \\
	&\quad+\frac{1}{453600 \pi^{3}}\left(\left(-46258 \epsilon^{4}+23337 \epsilon^{2}-71376\right) \pi^{2}\right. \notag \\
	&\quad\left.+\,107520 \epsilon^{4}+68096 \epsilon^{2}-175616\right) g^{6}
	+ \mathcal{O}\left(g^8\right) \,,
\end{align}
\begin{align}
	\frac{P_{3}^{\circ}}{\mathrm{i}} &= J_{3}^{\circ} \notag \\ 
	&= \sqrt{1-\epsilon^{2}}\, \left[-\frac{8 g^{3}}{35 \pi}+\frac{2 \left(68\epsilon^{2}+171\right) g^{5}}{945 \pi}\right. \notag \\
	&\quad-\frac{1}{4989600 \pi^{3}}\left(\left(449242 \epsilon^{4}+1015547 \epsilon^{2} -256536 \right)\pi^{2}\right. \notag \\
	&\quad\left.\left.-\,1520640\epsilon^{4}-4764672 \epsilon^{2}+10543104\right) g^{7}\right. \notag \\
	&\quad\left.+\,\mathcal{O}\left(g^9\right) \vphantom{ \frac{2 \left(68\epsilon^{2}+171\right) g^{5}}{945 \pi} }\right]\,,
\end{align}
\begin{align}
	Q_{2}^{\circ} &= E_{2}^{\circ} \notag \\ 
	&= \epsilon\, \left[ -\frac{4 g^{2}}{15 \pi}+\frac{4 \left(4\epsilon^{2}+9\right) g^{4}}{105 \pi}\right. \notag \\
	&\quad-\frac{1}{453600 \pi^{3}}\left(\left(36018 \epsilon^{4}-26537\epsilon^{2}+84816\right) \pi^{2}\right. \notag \\
	&\quad\left.\left.-\,107520 \epsilon^{4}-68096 \epsilon^{2}+175616\right) g^{6}
	+ \mathcal{O}\left(g^8\right) \vphantom{ \frac{4 \left(4\epsilon^{2}+9\right) g^{4}}{105 \pi} }\right]\,,
\end{align}
\begin{align}
	\frac{Q_{3}^{\circ}}{\mathrm{i}} &= B_{3}^{\circ} \notag \\ 
	&= \epsilon \sqrt{1-\epsilon^{2}}\, \left[-\frac{4 g^{3}}{35 \pi}+\frac{\left(62 \epsilon^{2}+51\right) g^{5}}{945 \pi}\right. \notag \\
	&\quad-\frac{1}{9979200 \pi^{3}}\left(\left(394522 \epsilon^{4}+54107 \epsilon^{2}-2568216\right) \pi^{2}\right. \notag \\
	&\quad\left.\left.-\,1520640 \epsilon^{4}-9022464 \epsilon^{2}+19058688\right) g^{7}\right. \notag \\
	&\quad\left.+\,\mathcal{O}\left(g^9\right) \vphantom{ \frac{\left(62 \epsilon^{2}+51\right) g^{5}}{945 \pi} }\right]\,.
\end{align}

The mass, angular momentum, electric and magnetic moments are alternating positive and negative, to be more precise,
\begin{align}
	&\text{for} \,\; l=0,2,4,...: \, & M_{2l}\geq0 \,, \quad\! J_{2l+1}\geq0 \,, \notag \\ 
	&                                               \, & E_{2l}\geq0 \,, \quad B_{2l+1}\geq0 \,, \\
	&\text{for} \,\; l=1,3,5,...: \, & M_{2l}\leq0 \,, \quad\! J_{2l+1}\leq0 \,, \notag \\
	&                                               \, & E_{2l}\leq0 \,, \quad\! B_{2l+1}\leq0 \,.
\end{align} 

It should be emphasized that the global prefactors, whereby
\begin{equation}
	J_{n} \sim \sqrt{1-\epsilon^{2}} \,, \quad E_{n} \sim \epsilon \,, \quad B_{n} \sim  \epsilon \sqrt{1-\epsilon^{2}} \,,
\end{equation}
ensure that for vanishing charge, i.e.\ $\epsilon=0$, all electromagnetic multipole moments $Q_{n}$ become zero and for vanishing rotation, i.e.\ $\epsilon=1$, all angular momentum and magnetic moments, $J_{n}$ and $B_{n}$, respectively.\footnote{Additionally, mass and electric moments contain only even powers of $g$ and angular momentum and magnetic moments only odd, due to a symmetric and antisymmetric transformation behaviour, respectively, under a change of sense of rotation. For further details, see \cite{Palenta_2013} or \cite{Rumler}.}

Note that disc configurations with $\epsilon=0$ rotate with maximal angular velocity and ones with $\epsilon=1$ have no rotation at all. This becomes obvious in the Newtonian limit where each dust particle in the disc is in an equilibrium of gravitational, electric and centrifugal force.

\begin{figure}[htb]
	\centering
	\includegraphics[width=0.48\textwidth]{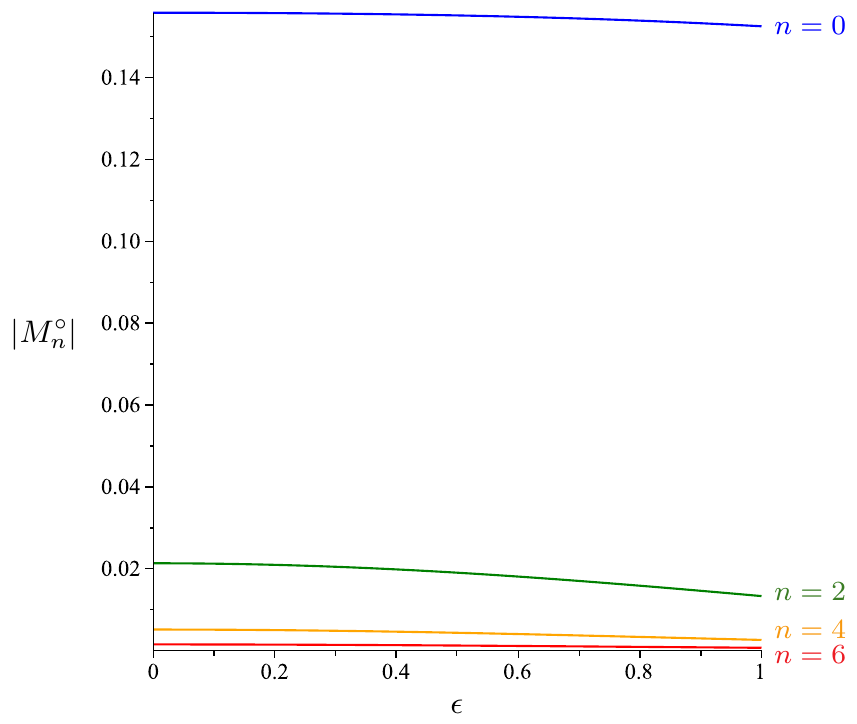}
	\caption{Normalized mass moments $M_{n}^{\circ}$ for $n=\{0,2,4,6\}$ as functions of the specific charge $\epsilon$ plotted for $g=0.6$.}
	\label{fig:1}
\end{figure}

\begin{figure}[htb]
	\centering
	\includegraphics[width=0.43\textwidth]{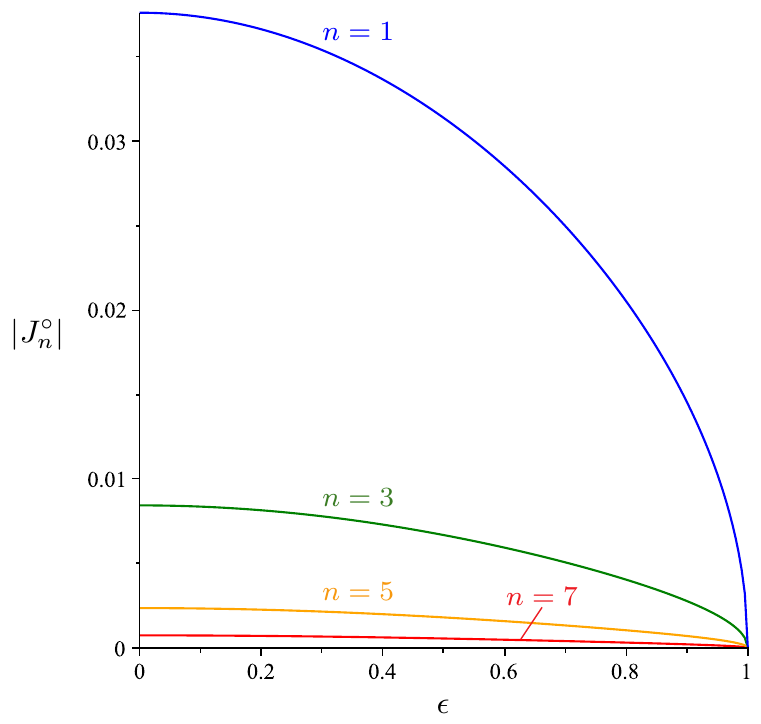}
	\caption{Normalized angular momentum moments $J_{n}^{\circ}$ for \mbox{$n=\{1,3,5,7\}$} as functions of the specific charge $\epsilon$ plotted for $g=0.6$.}
	\label{fig:2}
\end{figure}

\begin{figure}[htb]
	\centering
	\includegraphics[width=0.48\textwidth]{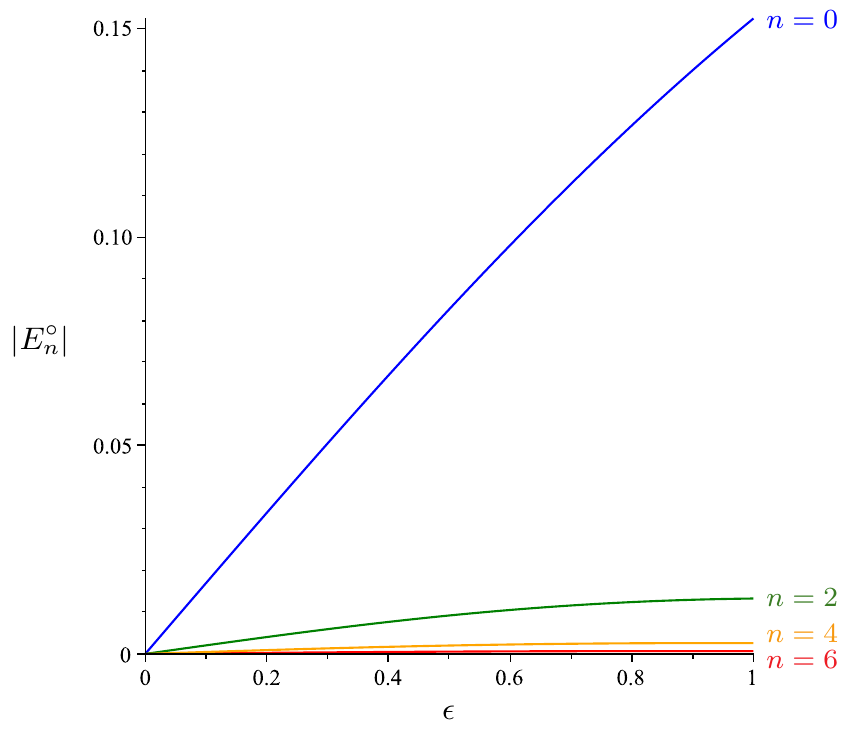}
	\caption{Normalized electric moments $E_{n}^{\circ}$ for $n=\{0,2,4,6\}$ as functions of the specific charge $\epsilon$ plotted for $g=0.6$.}
	\label{fig:3}
\end{figure}

\begin{figure}[htb]
	\centering
	\includegraphics[width=0.43\textwidth]{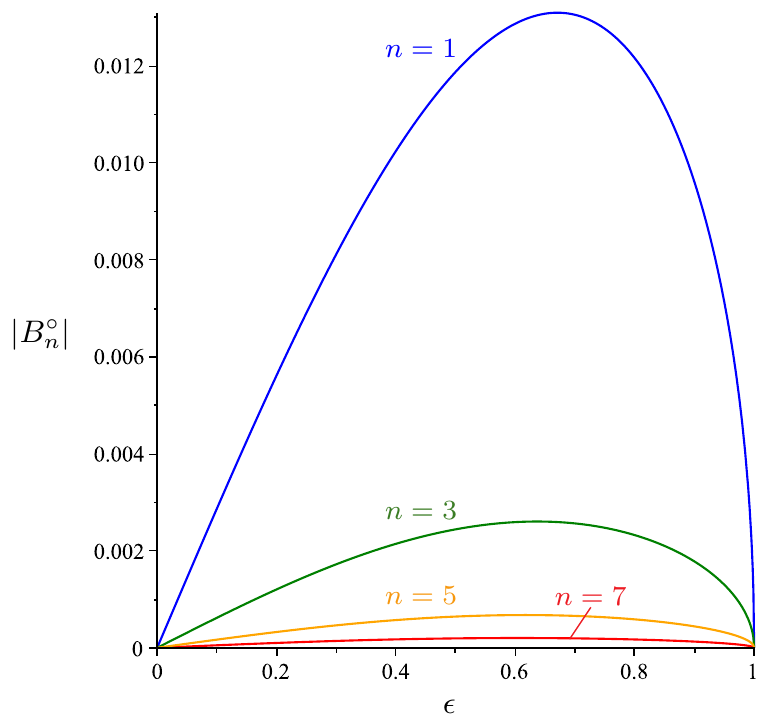}
	\caption{Normalized magnetic moments $B_{n}^{\circ}$ for $n=\{1,3,5,7\}$ as functions of the specific charge $\epsilon$ plotted for $g=0.6$.}
	\label{fig:4}
\end{figure}

The influence of the global prefactors is also reflected in \cref{fig:1,fig:2,fig:3,fig:4}. There, the dependence of the multipole moments normalized by the proper radius, $M_{n}^{\circ}$, $J_{n}^{\circ}$, $E_{n}^{\circ}$ and $B_{n}^{\circ}$, on the specific charge $\epsilon$ is depicted for \mbox{$n=0,...,7$}. The relativity parameter is set to $g=0.6$.


\begin{figure}[tb]
	\centering
	\includegraphics[width=0.48\textwidth]{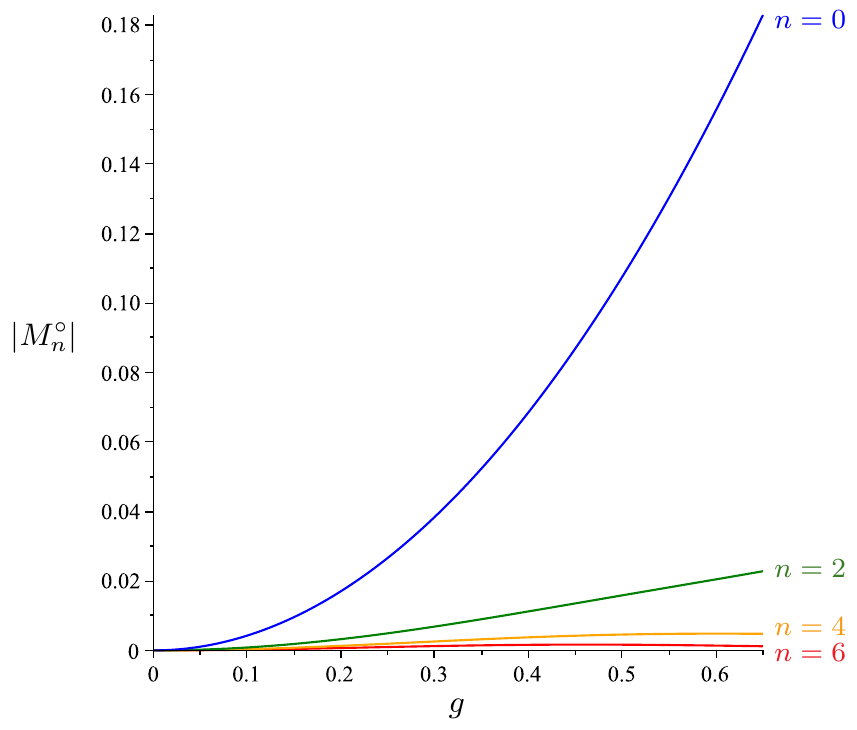}
	\caption{Normalized mass moments $M_{n}^{\circ}$ for $n=\{0,2,4,6\}$ as functions of the relativity parameter $g$ plotted for $\epsilon=0.3$.}
	\label{fig:1b}
\end{figure}
\begin{figure}[tb]
	\centering
	\includegraphics[width=0.48\textwidth]{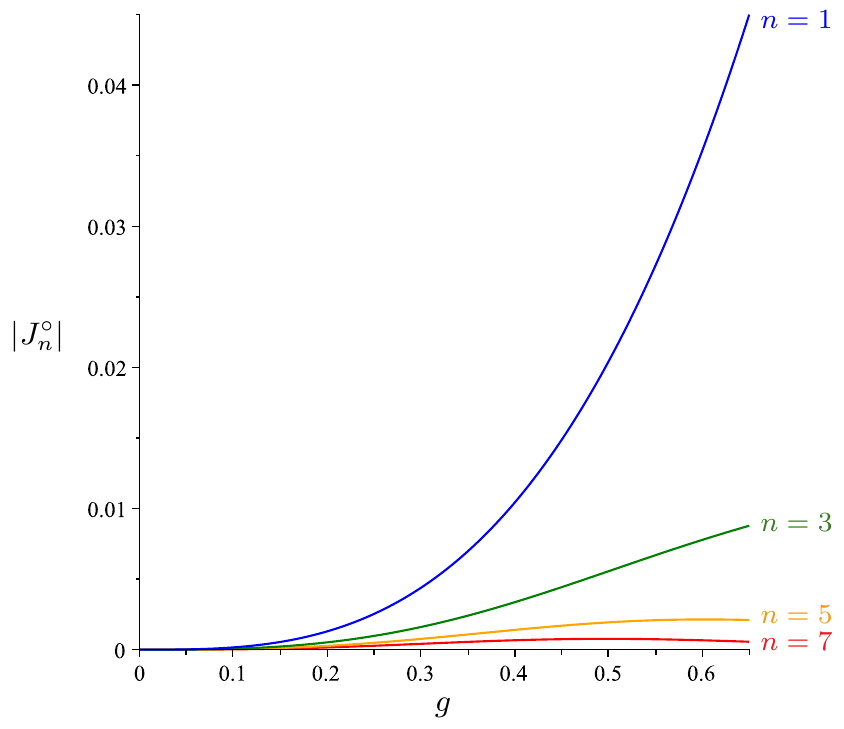}
	\caption{Normalized angular momentum moments $J_{n}^{\circ}$ for \mbox{$n=\{1,3,5,7\}$} as functions of the relativity parameter $g$ plotted for $\epsilon=0.3$.}
	\label{fig:2b}
\end{figure}

Figs.\ \ref{fig:1b} and \ref{fig:2b} show the dependence of the normalized gravitational multipole moments on the relativity parameter $g$, where $\epsilon=0.3$. As can be seen there, in the limit $g\to 0$ all gravitational multipole moments $P_{n}^{\circ}$ vanish 
and  with increasing $g$ the first ones (to be more precise, $n\leq 1$ for all $\epsilon$) grow monotonically.
The electromagnetic multipole moments $Q_{n}$ show a completely analogous behaviour.

Remarkably, as can be seen in \cref{fig:1,fig:2,fig:3,fig:4,fig:1b,fig:2b}, all multipole moments are perfectly ordered:
\begin{align}
	\left\vert M_{n} \right\vert &\geq \left\vert M_{n+2} \right\vert \,, \\
	\left\vert J_{n} \right\vert &\geq \left\vert J_{n+2} \right\vert \,, \\
	\left\vert B_{n} \right\vert &\geq \left\vert B_{n+2} \right\vert \,, \\
	\left\vert E_{n} \right\vert &\geq \left\vert E_{n+2} \right\vert \,.
\end{align}

Notice that this ordering as well as the alternating sign can also be observed for the (exact) gravitational multipole moments of the uncharged disc of dust \cite{Kleinwächter}.

\begin{figure}[htb]
	\centering
	\includegraphics[width=0.48\textwidth]{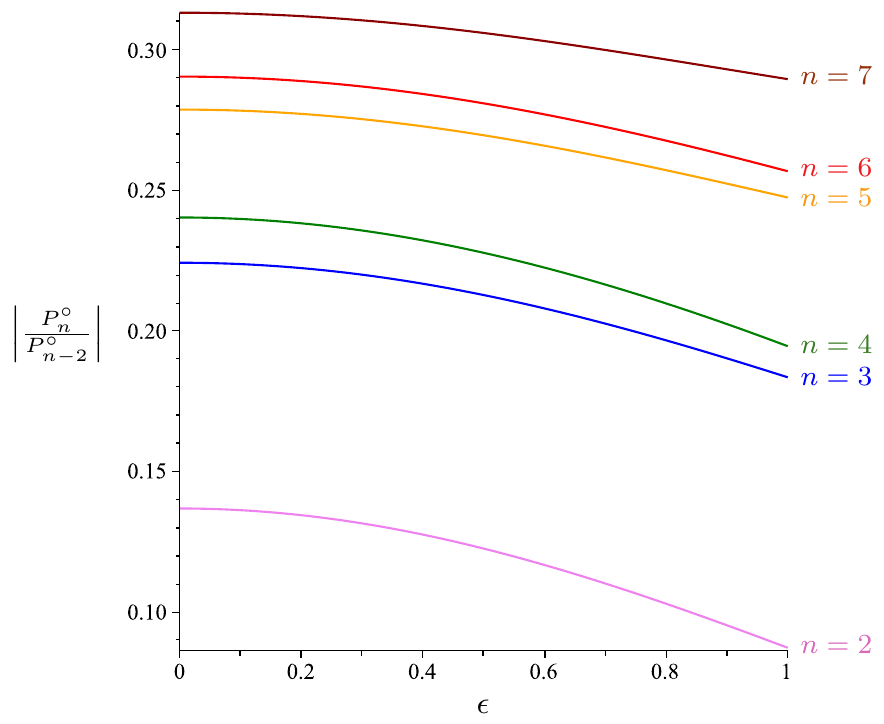}
	\caption{Ratios $\frac{P_{n}^{\circ}}{P_{n-2}^{\circ}}$ for $n=\{2,3,...,7\}$ as functions of the specific charge $\epsilon$ plotted for $g=0.6$.}
	\label{fig:ratio}
\end{figure}

\begin{figure}[htb]
	\centering
	\includegraphics[width=0.48\textwidth]{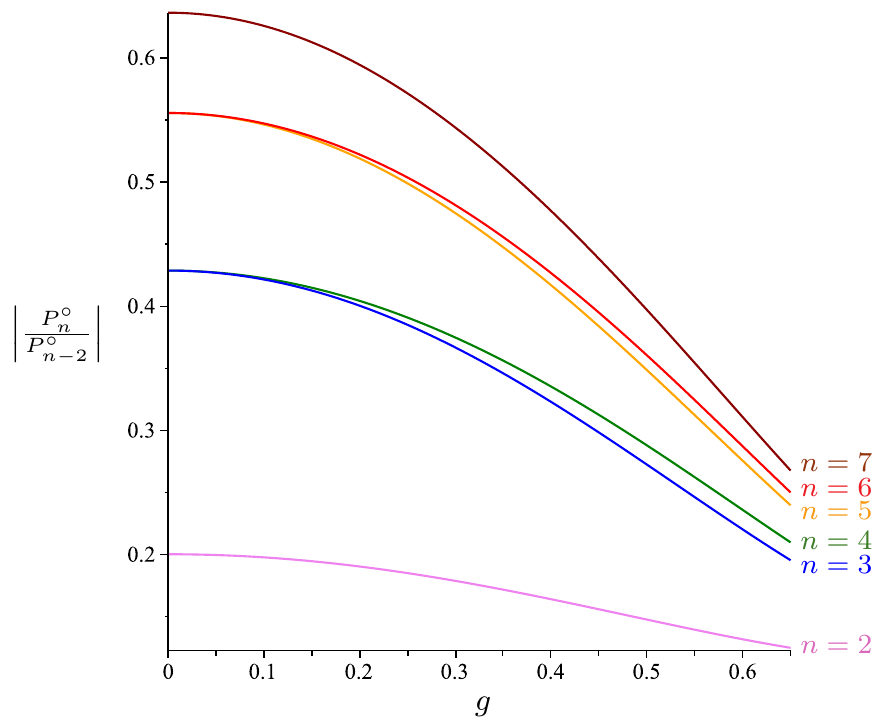}
	\caption{Ratios $\frac{P_{n}^{\circ}}{P_{n-2}^{\circ}}$ for $n=\{2,3,...,7\}$ as functions of the relativity parameter $g$ plotted for $\epsilon=0.3$.}
	\label{fig:ratio2}
\end{figure}

Interesting are also the ratios $\frac{P_{n}^{\circ}}{P_{n-2}^{\circ}}$ and $\frac{Q_{n}^{\circ}}{Q_{n-2}^{\circ}}$ of the individual moments. According to \cref{fig:ratio,fig:ratio2},  the absolute values of the ratios are greater, the higher the gravitational multipole moments are. The corresponding plots of the ratios of the electromagnetic multipole moments look almost identical to those of the gravitational moments and in fact agree at $g=0$.\footnote{This is not a coincidence, as in the Newtonian limit $\Lambda = \epsilon\, \Xi$ holds.} In addition, there is a pairwise agreement at $g=0$:  $\frac{P_{3}^{\circ}}{P_{1}^{\circ}} = \frac{P_{4}^{\circ}}{P_{2}^{\circ}} = -\frac{3}{7} \left(= \frac{Q_{3}^{\circ}}{Q_{1}^{\circ}} = \frac{Q_{4}^{\circ}}{Q_{2}^{\circ}}\right)$ and $\frac{P_{5}^{\circ}}{P_{3}^{\circ}} = \frac{P_{6}^{\circ}}{P_{4}^{\circ}} = -\frac{5}{9} \left(= \frac{Q_{5}^{\circ}}{Q_{3}^{\circ}} = \frac{Q_{6}^{\circ}}{Q_{4}^{\circ}}\right)$.


The decision to restrict our discussion to the first eight multipole moments $P_{n}$ and $Q_{n}$ and the relativity parameter to \mbox{$g\leq 0.65$} is based on the convergence behaviour of the multipole moments. As an example, to estimate the convergence of the electric moments, we calculate the resulting change by adding the $K$th order to the post-Newtonian expansion at order $K-1$ relative to the solution at $K$th order: 
\begin{align}\label{eq:conv}
	\delta E_{n\,\vert\,K}^{\circ} \coloneqq \frac{\left\vert E_{n\,\vert\,K}^{\circ} - E_{n\,\vert\,K-1}^{\circ}\right\vert}{\left\vert E_{n\,\vert\,K}^{\circ}\right\vert} \,, \\
 	\text{where} \quad E_{n\,\vert\,K}^{\circ} \coloneqq \sum_{k=1}^{K}E_{n2k}^{\circ}g^{2k} \,.
\end{align}
\begin{figure}[htb]
	\centering
	\includegraphics[width=0.48\textwidth]{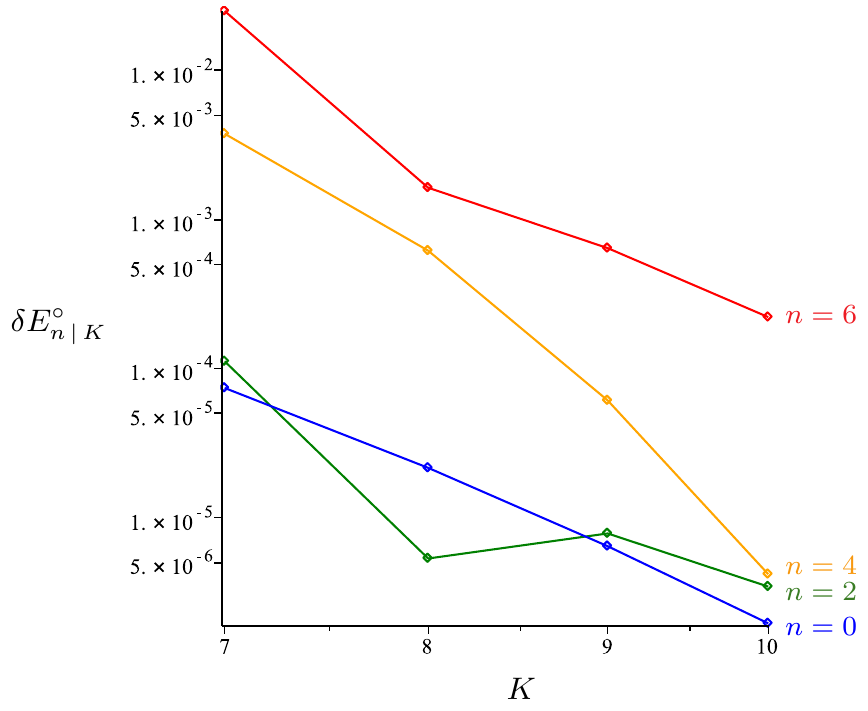}
	\caption{Convergence estimate of the normalized multipole moments $E_{0}^{\circ}$, $E_{2}^{\circ}$, $E_{4}^{\circ}$ and $E_{6}^{\circ}$ at order $k=K$ plotted for $g=0.65$ and $\epsilon=0$.}
	\label{fig:8}
\end{figure}
According to \cref{fig:8}, the relative change by adding the last order is for all $E_{n}^{\circ}$ ($n=0,..,6$) less than or equal to $\num{2.2e-4}$ for $g=0.65$. (In this case $\epsilon=0$ shows the worst convergence behaviour.) Moreover, as can already  be guessed from the calculation scheme, the higher the multipole moments, the less good the convergence.
For the mass, angular momentum and magnetic moments we get an even better, but otherwise analogous convergence behaviour.\footnote{For $\epsilon=0$ and $g=0.65$ one gets: $\delta M_{6\,\vert\,10}^{\circ}=\num{7.5e-5}$, $\delta J_{7\,\vert\,10}^{\circ}=\num{1.9e-4}$, $\delta B_{7\,\vert\,10}^{\circ}=\num{9.7e-5}$.}

\begin{table}[htb]
\centering
\begin{minipage}{8.35cm}
\caption{Normalized multipole moments $M_{6}^{\circ}$, $J_{7}^{\circ}$: charged disc in the limit $\epsilon=0$ (post-Newtonian expansion up to tenth order) versus uncharged disc (exact solution).}\label{tab1}
\begin{tabular*}{\textwidth}{@{\extracolsep{\fill}}lcccccc@{\extracolsep{\fill}}}
	\toprule
	&		&	$\text{charged disc}\rvert_{\epsilon=0}$	&	uncharged disc \\
	\midrule
	$g=0.55$	& 	$\left\vert M_{6}^{\circ}\right\vert$	&	$\num{1.6286550e-3}$	&	$\num{1.6286551e-3}$   \\		
	& 	$\left\vert J_{7}^{\circ}\right\vert$	&	$\num{8.0663923e-4}$	&	 $\num{8.0663952e-4}$	\\		
	\midrule
	$g=0.6$	& 	$\left\vert M_{6}^{\circ}\right\vert$	&	$\num{1.4875997e-3}$	&	$\num{1.4876000e-3}$  \\		
	& 	$\left\vert J_{7}^{\circ}\right\vert$	&	$\num{7.3636251e-4}$	& 	$\num{7.3636450e-4}$  \\		
	\midrule
	$g=0.65$	& 	$\left\vert M_{6}^{\circ}\right\vert$	&	$\num{1.3141025e-3}$	&	$\num{1.3141033e-3}$  \\		
	& 	$\left\vert J_{7}^{\circ}\right\vert$	&	$\num{6.3590588e-4}$	& 	$\num{6.3591807e-4}$  \\		
	\bottomrule
\end{tabular*}
\end{minipage}
\end{table}

A direct comparison of the derived multipole moments (in terms of the post-Newtonian expansion) in the limit $\epsilon = 0$ with the exact solutions of the uncharged rotating disc of dust \cite{Kleinwächter} shows an excellent agreement.
In \cref{tab1} we compare the sixth and seventh gravitational multipole moments for different values of the relativity parameter.
In particular, the relative error of the multipole moments of the charged rotating disc of dust in the limit $\epsilon=0$ with respect to the exact solutions at $g=0.65$ is $\num{6.1e-7}$ for $M_{6}^{\circ}$ and $\num{1.9e-5}$ for $J_{7}^{\circ}$.

\section{Multipole conjecture}
\label{sec:conjectioure}

Filter and Kleinwächter \cite{Filter} formulated an interesting conjecture about the multipole moments of (uncharged) rigidly rotating perfect fluid bodies: The absolute values of all higher multipole moments $P_{n}$ ($n\geq 2$) of an axistationary, rigidly rotating, perfect fluid body, surrounded by vacuum, are always greater than those of the corresponding moments of the Kerr spacetime with the same mass and angular momentum.

An obvious question now is whether this conjecture can be extended to more general, particularly charged, bodies.

The multipole moments of the Kerr-Newman spacetime, with mass $M^{\text{KN}}$, angular momentum $J^{\text{KN}}$ and charge $Q^{\text{KN}}$, are given by \cite{Sotiriou_2004}:
\begin{align}
&P_{n}^{\text{KN}} = m_{n}^{\text{KN}} = M^{\text{KN}}\left(\mathrm{i}\frac{J^{\text{KN}}}{M^{\text{KN}}}\right)^{n} \,, \\
&Q_{n}^{\text{KN}} = q_{n}^{\text{KN}} = Q^{\text{KN}}\left(\mathrm{i}\frac{J^{\text{KN}}}{M^{\text{KN}}}\right)^{n} \,.
\end{align}
With these we state:
\begin{theorem}
For the gravitational and electromagnetic multipole moments, $P_{n}$ and $Q_{n}$, of an isolated, axisymmetric, stationary, physically well-defined body of ordinary matter, with mass $M$, angular momentum $J$ and charge $Q$, holds for all $n \geq 2$:
\begin{equation}
\left\vert P_{n} \right\vert \geq \left\vert \frac{J^{n}}{M^{n-1}}\right\vert \,, \quad
\left\vert Q_{n} \right\vert \geq \left\vert \frac{QJ^{n}}{M^{n}}\right\vert \,.
\end{equation}
Furthermore, in case of $J\neq 0$, equalities apply if and only if the body reaches a black hole limit.
\end{theorem}





The goal of this section is to test the generalized multipole conjecture using the charged rotating disc of dust as a concrete and physically meaningful candidate that satisfies the requirements of the conjecture.\footnote{Note that the Kerr-Newman spacetime furnished with the mass $M$, angular momentum $J$ and charge $Q$ of the charged rotating disc of dust does not describe a black hole but a hyperextreme solution ($Q^2  + \frac{J^2}{M^2} > M^2$).}
To this end we plot the quantities $X_{n}$ and $Y_{n}$ that have to be less than or equal to $1$ if the conjecture is valid:
\begin{equation}
X_{n} \coloneqq \left\vert \frac{J^{n}}{M^{n-1}P_{n}}\right\vert \leq 1 \,, \quad
Y_{n} \coloneqq \left\vert \frac{QJ^{n}}{M^{n}Q_{n}}\right\vert \leq 1 \,.
\end{equation}

\begin{figure}[htb]
\centering
\includegraphics[width=0.43\textwidth]{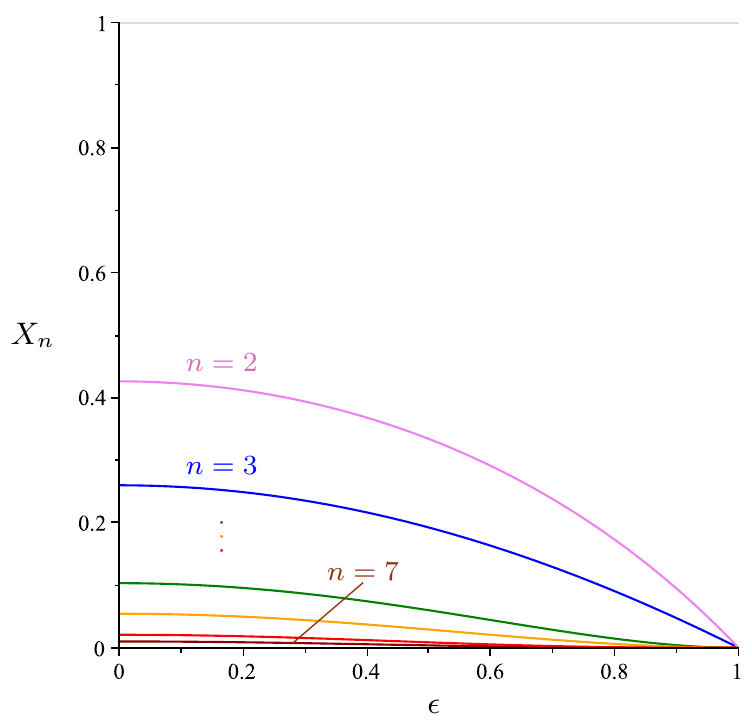}
\caption{$X_{n}$ for $n=\{2,3,...,7\}$ as functions of the specific charge $\epsilon$ plotted for $g=0.6$.}
\label{fig:9}
\end{figure}

\begin{figure}[htb]
\centering
\includegraphics[width=0.43\textwidth]{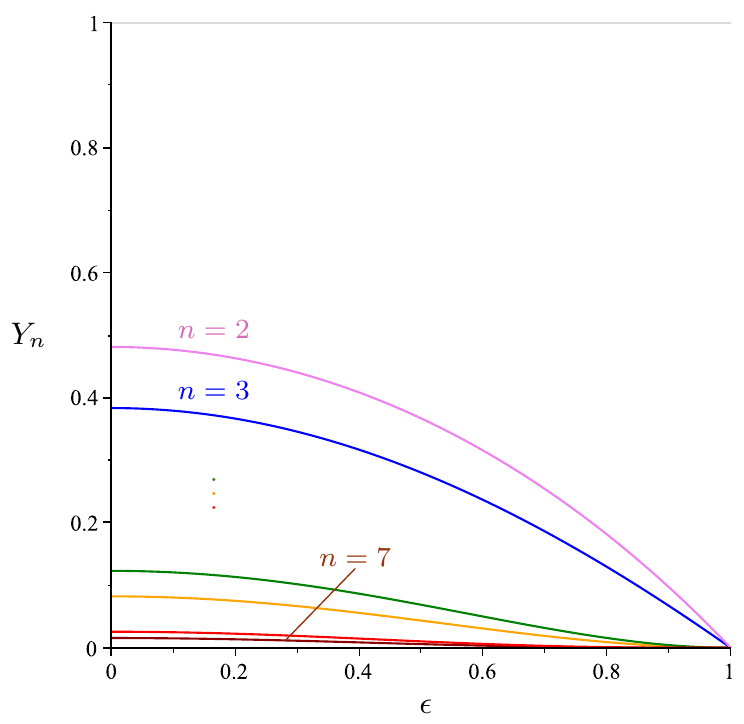}
\caption{$Y_{n}$ for $n=\{2,3,...,7\}$ as functions of the specific charge $\epsilon$ plotted for $g=0.6$.}
\label{fig:9b}
\end{figure}

Notice that due to the global prefactors of $P_{n}$ and $Q_{n}$ the quantities $X_{n}$ and $Y_{n}$ stay regular in the limits $\epsilon \to 1$ and $\epsilon \to 0$. In fact, $X_{n}$ and $Y_{n}$ vanish for $\epsilon \to 1$ and take finite values in the limit $\epsilon \to 0$ (see also \cref{fig:9,fig:9b}).

In the limit $\epsilon\to1$ the angular momentum $J$ vanishes and therefore also all Kerr-Newman multipole moments, $P_{n}^{\text{KN}}$ and $Q_{n}^{\text{KN}}$, with $n\geq1$ . Since the multipole moments of the disc, $P_{n}$ and $Q_{n}$, become zero only for odd, but not for even $n$, the disc spacetime clearly does not attain the Reissner-Nordström solution in this limit.

\begin{figure}[htb]
\centering
\includegraphics[width=0.48\textwidth]{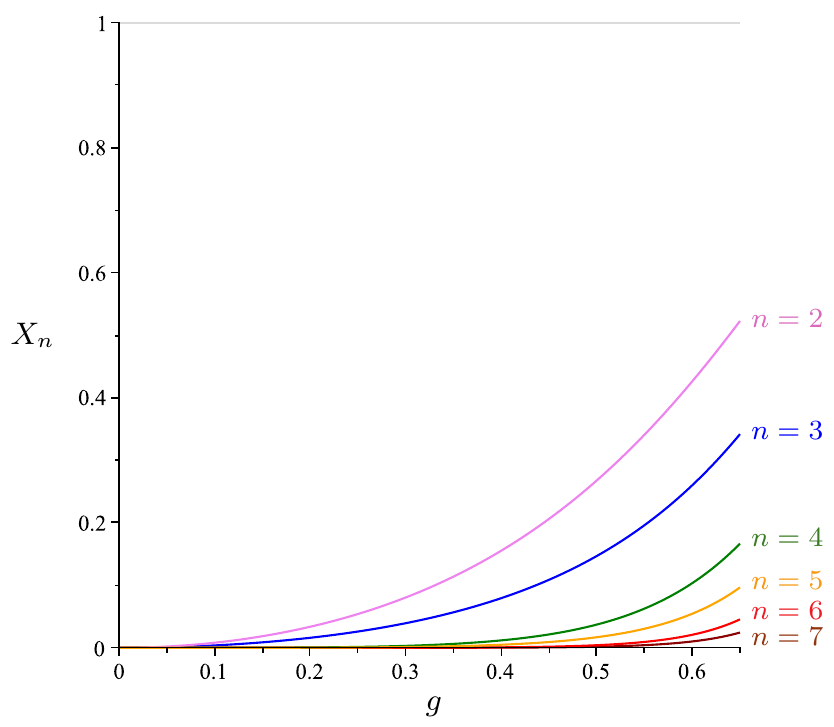}
\caption{$X_{n}$ for $n=\{2,3,...,7\}$ as functions of the relativity parameter $g$ plotted for $\epsilon=0$.}
\label{fig:5}
\end{figure}
\begin{figure}[htb]
\centering
\includegraphics[width=0.48\textwidth]{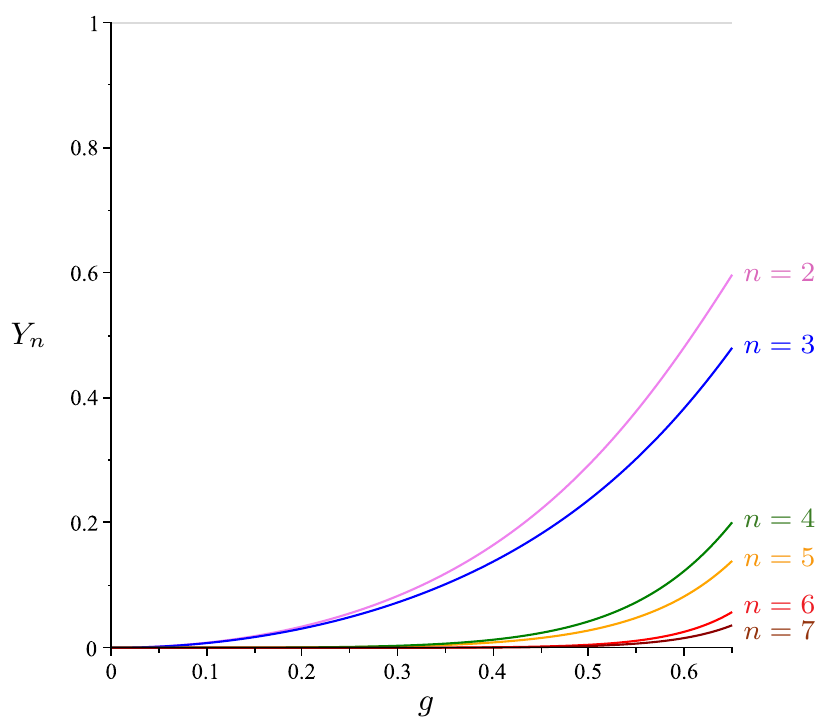}
\caption{$Y_{n}$ for $n=\{2,3,...,7\}$ as functions of the relativity parameter $g$ plotted for $\epsilon=0$.}
\label{fig:5b}
\end{figure}


As can be seen in \cref{fig:5,fig:5b} and \cref{fig:9,fig:9b}, indeed all $X_{n}$ and $Y_{n}$ are smaller than $1$ in the region $g\in\left[0,0.65\right]$. Furthermore, there is a similar ordering as for the individual moments:
\begin{equation}
X_{n} \geq X_{n+1} \,, \quad Y_{n} \geq Y_{n+1} \,.
\end{equation}
Therefore, most critical are $X_{2}$ and $Y_{2}$ and for increasing $n$ the conjecture is even better fulfilled. 

On the other hand, the first $X_{n}$ and $Y_{n}$ are also the quantities with the best convergence behaviour. As $n$ becomes larger, the convergence of $X_{n}$ and $Y_{n}$ (as it contains $P_{n}$ and $Q_{n}$, respectively) becomes less and less good. The relative change (as defined in \cref{eq:conv}) by adding the last order is $\delta X_{7\,\vert 10}=\num{1.9e-2}$ and $\delta Y_{7\,\vert 10}=\num{9.7e-3}$, where $g=0.65$ and $\epsilon=0$.\footnote{Specifically for $X_{7}$ and $Y_{7}$, the convergence gets slightly worse for higher $\epsilon$: $\delta X_{7\,\vert 10}=\num{4.1e-2}$ and $\delta Y_{7\,\vert 10}=\num{3.4e-2}$, for $\epsilon=1$ and $g=0.65$. However, as $X_{n}$ and $Y_{n}$ for high $n$ are not decisive for the verification of the conjecture, this is not problematic.}

The good convergence of $X_{2}$ and $Y_{2}$ allows verification of the conjecture for slightly higher values of the relativity parameter until $g \approx 0.8$ and $g \approx 0.75$, respectively.\footnote{For $\epsilon=0$: $\delta X_{2\,\vert 10}=\num{1.9e-2}$ at $g = 0.8$ and $\delta Y_{2\,\vert 10}=\num{2.8e-2}$ at $g = 0.75$.} The conjecture remains true there.
For the limit $g \to 1$, Breithaupt et al.\ \cite{Breithaupt_2015} provided strong evidence that the multipole moments of the charged rotating disc of dust converge to those of the (extreme) Kerr-Newman spacetime.

Additionally, one can observe from \cref{fig:9,fig:9b} that all values of $X_{n}$ and $Y_{n}$ decrease equally with increasing $\epsilon$. This fact is very convenient for our purposes, since for the most critical case, $\epsilon=0$, the exact multipole moments \cite{Kleinwächter} of the uncharged disc of dust \cite{Neugebauer:1993ct, PhysRevLett.75.3046, RFE} are available. 
The relative error of $X_{2}\vert_{\epsilon=0}$ with respect to the corresponding exact value of the uncharged disc of dust is $\num{1.3e-2}$ for $g=0.8$.
For the gravitational multipole moments of the uncharged disc of dust Filter and Kleinwächter showed that the conjecture is fulfilled \cite{Filter}.

In summary, the conjecture holds at least up to $g\approx0.8$ for the gravitational and up to $g\approx0.75$ for the electromagnetic multipole moments. In the limit $g \to 1$ the quantities $X_{n}$ and $Y_{n}$ converge to $1$ according to \mbox{Breithaupt} et al.\ and for the most critical case, \mbox{$\epsilon=0$}, by means of the exact solution of the uncharged disc of dust it was proven that $X_{n} \leq 1$. All this taken together, we conclude that (within the scope of accuracy of the post-Newtonian expansion) the generalized multipole conjecture stated above is fulfilled for the multipole moments of the charged rotating disc of dust.

\section{Conclusions and outlook}
\label{sec:conclusions}

The individual mass, angular momentum, electric and magnetic moments of the charged rotating disc of dust, in agreement with the uncharged disc \cite{Kleinwächter}, follow an ordering in which the higher they are, the less they contribute to the gravitational and electromagnetic field. This effect is in addition to the suppression of higher orders in the asymptotic expansions. 
Therefore, primarily the first multipole moments are relevant for the far-field behaviour of the gravitational and electromagnetic field of the charged rotating disc of dust.


Of course, it would be desirable to have a proof of the generalized multipole conjecture. This would first require a precise notion of physical well-definedness.\footnote{Perhaps one should consider a stronger condition, like uniformly rotating, charged perfect fluid bodies with convective electric currents only, see \cite{Novak_2003}.}
In case of non-vanishing angular momentum the conjecture could serve as a powerful tool to distinguish ordinary physical bodies from black holes by measuring their multipole moments in the far-field. In fact, if the necessary condition $Q^2 + \frac{J^2}{M^2} \leq M^2$ for a black hole is satisfied, the measurement of the multipole moments for $n = 2$ would already make the difference.

\begin{acknowledgments}
This work has been funded by the Deutsche
Forschungsgemeinschaft (DFG) under Grant No.
406116891 within the Research Training Group RTG
2522/1. 
The authors would like to thank Andreas Kleinwächter for the computed multipole moment values of the uncharged disc of dust
and \mbox{Martin} Breithaupt for the provided results of the post-Newtonian expansion up to tenth order.
\end{acknowledgments}



\bibliography{moments-bibliography}

\end{document}